\documentclass{vgtc}                          % final (conference style)
\graphicspath{{figures/}{pictures/}{images/}{./}} % where to search for the images

\usepackage{times}                     % we use Times as the main font
\usepackage{tabu}                      % only used for the table example
\usepackage{booktabs}                  % only used for the table example
\usepackage{lipsum}                    % used to generate placeholder text
\usepackage{mwe}                       % used to generate placeholder figures

\usepackage{mathptmx}                  % use matching math font
\usepackage{amsmath}                   % for \text in math mode
\usepackage{amssymb}
\usepackage{graphicx}
\usepackage{multirow}
\usepackage{svg}
\usepackage[normalem]{ulem}

\onlineid{0}

\vgtccategory{Research}

\vgtcinsertpkg

\title{Cognitive State Inference from VR Motion via Motion Foundation Model}

\author{Kaiang Wen\thanks{e-mail: kwen2@hawk.illinoistech.edu} %
\and Mark Roman Miller\thanks{e-mail: mmiller30@illinoistech.edu}} %
\affiliation{\scriptsize Department of Computer Science \\ Illinois Institute of Technology}

\teaser{
  \centering
      \includegraphics[width=\linewidth,
                     trim=0 0 0 0,
                     clip]{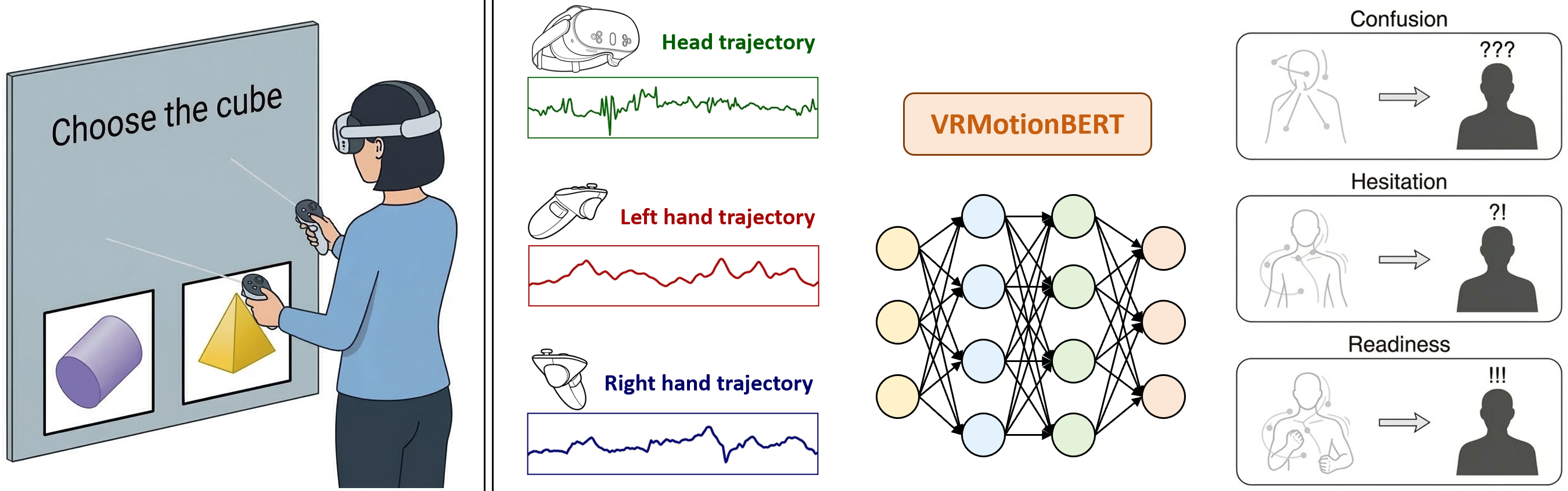}
  \caption{
\textbf{Overview of cognitive-state inference from sparse VR motion.}
\textbf{Left:} VR Experiment. Participants answer two-choice conceptual questions in VR, designed to elicit transient decision-related states such as \textit{confusion}, \textit{hesitation}, and \textit{readiness}. 
\textbf{Middle:} Only sparse head and hand trajectories (6DoF) are recorded as input signals. 
\textbf{Right:} VRMotionBERT, a VR-native motion foundation model, infers fine-grained cognitive states directly from these trajectories. 
}
  \label{fig:teaser}
}

\abstract{

As virtual reality (VR) becomes widespread, head and hand motion data captured by consumer systems has become substantially more common. However, the extent of what can be inferred from such motion remains unclear. This paper investigates whether \textit{transient cognitive states}, specifically \textit{confusion, hesitation, and readiness} during different stages of decision-making, can be inferred from VR telemetry alone.
We introduce a novel dataset of head and hand motion collected during structured decision-making tasks, with frame-level annotations of these states. We evaluate classical machine learning models, temporal neural networks, and motion foundation models under two protocols: (1) future-in-time prediction for the same users, and (2) cross-user generalization to unseen users. We further propose a VR-native motion adapter that maps sparse VR telemetry to representations compatible with motion foundation models pretrained on large-scale full-body motion data, enabling transfer without explicit full-body reconstruction. To our knowledge, this is the first work to adapt a motion foundation model to VR motion for a classification task.
Results show that motion-only sensing captures meaningful signals of cognitive states, and that pretrained motion foundation models generalize more effectively than classical and temporal models even with a small dataset of 24 participants. Our approach achieves 82\% accuracy, comparable to and sometimes surpassing human observers.  These findings suggest that VR motion encodes richer behavioral information than previously assumed and highlight the potential of large-scale motion pretraining for XR applications. We will release the dataset and modeling framework to support future research.
} % end of abstract


\begin{document}


% \firstsection{Introduction}

\maketitle

% Note to self: This is the future I imagined when I first started doing this project. If VR devices has the built-in cognitive state detection model running in the background, it can automatically tell when user is getting confused by a new App, and pop up the new user instruction again. When hesitation is detected when the user is going through a tutorial, a helping agent, whether robot or human, can come and offer to help. If readiness is detected, the system should know user has been prepared for the next step and can moved quickly to the next step and reduce waiting time. This is the fascinating future I imagined if our project is landed for partical use.

% \input{revision_plan}

% \textcolor{blue}{Kaiang: Editing Introduction finished (Mar 13)}
\section{Introduction}

% Virtual Reality (VR) is rapidly evolving from a specialized tool for simulation and entertainment into a mainstream computing platform for work, education, and social interaction. As users spend more time in these immersive environments, the quality of human-computer interaction becomes paramount. The next generation of VR systems must move beyond explicit, command-based interfaces and develop the capacity for implicit, nuanced understanding. This requires an ability to perceive and adapt to a user's cognitive state in real-time, creating experiences that are more intuitive, supportive, and effective. The key to unlocking this capability lies in decoding the rich, continuous, and often subconscious stream of motion data generated by every user.

Virtual Reality (VR) and Augmented Reality (AR) are rapidly evolving from a specialized tool for simulation and entertainment into a mainstream computing platform for work, education, and social interaction.
Consumer VR systems continuously track users' head and hand motion data, making such VR telemetry ubiquitous.
Prior work has shown that VR motion data can be highly identifying and reflect broad user traits, often aligning with intuitive "folk theories" of body language. However, it remains unclear to what extent motion kinematics encode more nuanced cognitive states, 
% such as confusion, hesitation, and readiness, 
which lack clear correlates with motion. 

% To investigate this, we introduce a novel dataset of head and hand motion with frame-level annotations of these states collected during structured decision-making tasks. 

In this paper, we investigate whether transient cognitive states can be inferred from VR telemetry alone. We focus on three states that arise during a decision-making process, \textit{confusion}, \textit{hesitation}, and \textit{readiness}. These states are of particular interest because they are highly relevant to immersive interaction, yet are subtle, temporally dynamic, and not directly observable and distinguishable from head and hand motion. If these cognitive states can be detected from standard VR telemetry, it will demonstrate that such motion contains rich information about users' thinking process, and extend our understanding of motion beyond overt physical behavior. 
% (I'm not quite sure about the above paragraph's writing)

A growing body of research shows that motion carries rich signatures of internal states. It is intuitive that movement can reveal broad traits: slow or reduced motion has been linked to depression \cite{wang2021detecting, gahalawat2023explainable}, while posture and gesture can signal archetypal emotions \cite{ouguz2024emotion}. These findings echo common “folk theories” of body language. Yet the ceiling for what motion can reveal is likely much higher than intuition suggests. For example, it is unsurprising that a face in a photograph can identify a person, but less obvious that the unique current drawn by household appliances can reveal which devices are active from a single outlet \cite{hart1992nonintrusive}. In VR, head and hand trajectories can be so distinctive that they act as biometric fingerprints \cite{miller2020personal}, identifying users among tens of thousands \cite{nair2023unique}. If motion can reveal who a user is, could it also reveal what they are thinking moment to moment?

The link between motion kinematics and the subtle, moment-to-moment cognitive states that lack clear physical correlates remains unexplored. This leads to our research questions: \textit{Is it possible to use solely head and hand motion to infer human's nuanced cognitive states, such as confusion, hesitation, and readiness? And how well can machine learning models perform on this task compared to human observers?}

Our work addresses this counterintuitive frontier by focusing on fine-grained cognitive states during user's decision-making: confusion, hesitation, and readiness, that lack clear, observable correlates in body language. While prior research has shown that motion can capture mood or general affect \cite{ouguz2024emotion}, and physiological sensing can capture workload or stress ~\cite{suzuki2024measuring, lobo2016cognitive}, no study has systematically tested whether head and hand motion captured by consumer VR system alone is indicative of these subtle, nuanced cognitive processes. 

% To investigate this problem, we introduce a new dataset of head and hand motion paired with frame-level self-annotations of reading, confusion, hesitation, and readiness, collected during structured decision-making tasks. We then evaluate whether machine learning models can reveal nuanced cognitive states from motion alone and compare their performance to that of human observers. This study provides the first systematic evidence that standard VR motion telemetry is identifying for nuanced cognitive states, opening the door to adaptive VR systems that anticipate and respond to users’ unspoken needs.

To investigate this problem, we introduce a novel dataset of VR head and hand motion collected during structured decision-making tasks, with frame-level annotations of confusion, hesitation, and readiness. We further propose a \textit{VR-native motion adapter} that projects sparse VR telemetry into a representation compatible with motion foundation models pretrained on large-scale full-body motion data, enabling knowledge transfer without requiring full-body pose reconstruction. We evaluate classical machine learning models, temporal neural networks, and motion foundation models under two protocols: \textit{(1) future-in-time prediction for the same users} and \textit{(2) cross-user generalization to previously unseen users}.

Our results show that cognitive states can be inferred from head and hand motion alone, suggesting that standard VR telemetry contains richer behavioral information than is typically assumed. Moreover, pretrained motion foundation models generalize more effectively to unseen users than classical machine learning models and temporal neural networks, despite the relatively small size of our dataset. Together, these findings demonstrate both the promise of motion-based cognitive-state inference in XR and a practical pathway for adapting large-scale motion representations to sparse VR tracking data.

% Our contributions can be summarized as follows:
% \begin{itemize}
%     \item We introduce a novel dataset of VR head and hand motion captured from consumer VR systems, with frame-level annotations of cognitive states including confusion, hesitation, and readiness. 
%     \item We propose a VR-native motion adapter that projects sparse VR telemetry into a representation compatible with motion foundation models pretrained on large-scale full-body motion datasets, enabling knowledge transfer without requiring full-body pose reconstruction.
%     \item We conducted extensive experiments to evaluate classical, convolutional, recurrent, and motion foundation models against a human baseline, identifying strengths and limits of each.
%     \item We provide the first systematic evidence that subtle cognitive states can be inferred from standard VR telemetry alone, suggesting that VR motion contains rich behavioral signals that extend beyond clearly physical movement analysis.
%     \item Our dataset and modeling framework will be made publicly available to support future research.
% \end{itemize}

Our contributions can be summarized as follows:
\begin{itemize}
    \item We introduce a novel dataset of VR head-and-hand motion captured from consumer VR systems, with frame-level annotations of cognitive states including \textit{confusion}, \textit{hesitation}, and \textit{readiness}.
    
    \item We propose a VR-native motion adapter that projects sparse VR telemetry into a representation compatible with motion foundation models pretrained on large-scale full-body motion datasets, enabling knowledge transfer without requiring full-body pose reconstruction.
    
    % \item We conduct extensive experiments comparing classical machine learning models, convolutional and recurrent neural networks, and motion foundation models against a human baseline, identifying the strengths and limitations of each.
    
    % \item We show that subtle cognitive states can be inferred from standard VR telemetry alone, suggesting that VR motion contains rich behavioral signals beyond clearly physical movement analysis.

    \item We show that subtle decision-related cognitive states
    % specifically \textit{confusion}, \textit{hesitation}, and \textit{readiness}, 
    can be detected and distinguished from sparse VR motion alone. This finding highlights the broader potential of VR telemetry as a rich behavioral signal, suggesting that it may support more ambitious forms of user-state inference than previously assumed.
    
    \item We find that pretrained motion foundation models generalize more effectively to previously unseen users than classical machine learning and temporal models, highlighting the value of large-scale motion pretraining for XR inference tasks.
    
    % \item Our dataset and modeling framework will be made publicly available to support future research.
\end{itemize}

\section{Related Work}

\subsection{Internal State Inference from Behavioral and Physiological Signals}

A long-standing goal in human-centered computing is to infer latent internal states, such as emotion, cognitive load, attention, anxiety, or decision-related processes. In XR in particular, prior approaches mostly rely on physiological or multi-modal sensing, including EEG~\cite{wang2026subject, johnson12026seeing, qu2025enhancing, mai2021affective}, heart rate~\cite{johnson12026seeing, sonautomated}, galvanic skin response~\cite{qu2025enhancing, wang2026subject}, gaze~\cite{johnson12026seeing, sonautomated}, or other bio-signals~\cite{mavridou2018towards, karmakar2023real}, to estimate users' internal states. While effective, these approaches typically require sensing channels beyond those available in standard consumer VR systems.

A complementary line of work studies whether internal states can be inferred from behavioral signals, including body motion. Prior work has shown that human emotions can be recognized from the combination of upper body movement combined with facial expressions~\cite{ilyas2021deep}, suggesting that motion may serve as a practical proxy for otherwise unobservable internal processes.
More recent work has further demonstrated that full-body movement can reveal stress, workload, and uncertainty~\cite{brunye2025inferring}, that fine-grained hand gestures can support affect and cognitive load recognition~\cite{chua2024motion}, and that specific factors of full-body posture and movement are associated with emotional expressions~\cite{mahfoudi2022emotion}.
% Chua et al. detect affect and cognitive load from fine-grained hand gestures~\cite{chua2024motion}.
% Mafoudi et al. identify motion factors that express emotions in full-body movement and posture~\cite{mahfoudi2022emotion}.

% In our work, we focus on three transient cognitive states that arise during decision-making: \textit{confusion}, \textit{hesitation}, and \textit{readiness}. We draw on the rational planning tradition, which describes decision-making as a sequence of stages including problem definition, identification of alternatives, and evaluation before action~\cite{simon1955behavioral, payne1993adaptive, pirolli1999information}. This perspective helps motivate why confusion and hesitation, though often conflated in everyday language, correspond to different phases of the decision process in our task design.

% Most prior XR work on internal-state inference has either emphasized physiological and multimodal sensing or required fine-grained full-body motion or hand gestures, whereas our work asks whether transient decision-related cognitive states can be inferred from standard VR head-and-hand motion alone.

Together, these findings suggest that behavioral motion signals can encode meaningful information about internal states. However, most prior XR work has either emphasized physiological and multi-modal sensing or relied on fine-grained full-body motion or hand-gesture data. In contrast, our work asks whether transient cognitive states can be inferred from the sparse head\&hand motion alone, which is already available in standard consumer VR systems.

\subsection{What Can Be Inferred from VR Head\&Hand Motion?}

Standard VR telemetry is increasingly recognized as a rich behavioral signal. Prior work has shown that head and hand motion alone can strongly reveal users' identity. For example, VR head\&hand Motion has been shown to be personally identifying even in ordinary viewing scenarios~\cite{miller2020personal}, and large-scale studies have demonstrated robust user identification from head\&hand motion alone~\cite{nair2023unique}. Other work has shown that VR motion can reveal broader personal attributes beyond identity, further underscoring the informativeness of the sparse head\&hand motion data~\cite{nair2023inferring}.

These findings suggest that VR motion contains much richer information than expected. At the same time, prior work in VR motion analysis has often focused on identity, user attributes, locomotion, interaction behavior, or other outcomes closely tied to observable movement patterns~\cite{serpush2022wearable, sandeep2020application, zeng2025predicting, zhang2018human, zhang2024hybrid}. Existing studies also commonly rely on classical machine learning pipelines with handcrafted features or standard temporal models, including methods such as SVMs~\cite{hearst1998support}, tree-based ensembles~\cite{breiman2001random}, and recurrent or convolutional networks~\cite{sherstinsky2020fundamentals, o2015introduction}.

Closer to our setting, recent work has begun to explore affective or cognitive states from VR motion, suggesting that motion-only inference of internal states is an emerging and promising direction~\cite{chua2024motion, ouguz2024emotion}. However, to our knowledge, prior work has not systematically examined the joint prediction of distinct decision-related cognitive states, specifically \textit{confusion}, \textit{hesitation}, from standard VR telemetry. 
In our setting, models are required not only to detect cognitive-state signals from motion, but also to distinguish subtle differences between states such as \textit{confusion} and \textit{hesitation}.
Moreover, it remains unclear whether motion foundation models pretrained on large-scale full-body motion datasets can transfer effectively to sparse head\&hand motion.

\subsection{Motion Foundation Models and Sparse-to-Full-Body Transfer}

Recent work has shown growing interest in large-scale pretrained models for human motion, learned from increasingly large and diverse full-body motion datasets~\cite{joo2015panoptic, mahmood2019amass, punnakkal2021babel, guo2022generating, wang2024scaling, lin2025quest}. 
Many of these large motion models are multi-modal, incorporating both motion and language, and are primarily developed for tasks such as motion generation, editing, and cross-modal synthesis~\cite{jiang2023motiongpt, baharani2025mofm, li2025unimotion}. 
MotionBERT~\cite{zhu2023motionbert}, by contrast, is pretrained on motion alone and shows that pretrained motion encoders can transfer effectively to downstream motion understanding tasks such as human pose estimation and action recognition.
% This makes it better aligned with our problem, where the input consists solely of sparse VR head-and-hand motion and the goal is inference rather than generation.
% Therefore, we adopt MotionBERT as the main motion foundation model for our tasks
These developments raise important questions for XR: 
% \textit{can motion representations learned from large-scale full-body motion datasets benefit inference from sparse VR telemetry?}
Can pretrained motion foundation models benefit internal state inference from sparse VR motion? More specifically, it remains unclear whether pretrained motion foundation models can understand the subtle differences between cognitive states like confusion, hesitation, and readiness, and whether they improve generalization to previously unseen users compared with classical machine learning, temporal, and convolutional models.

% To adopt motion foundation models in VR motion, a central challenge is the input dimension mismatch between the two domains.
A central challenge in applying motion foundation models to VR motion is the representational mismatch between the two domains.
Motion foundation models are typically trained on full-body pose sequences with over 17 tracked joints, while consumer VR systems provide only head and hand trajectories. One possible solution is to reconstruct full-body pose from sparse VR telemetry. Prior work such as AvatarPoser reconstructs dense full-body poses from sparse VR head\&hand motion~\cite{jiang2022avatarposer, zheng2023realistic, castillo2023bodiffusion}, providing a plausible route for adapting sparse VR input to full-body motion models. Recent work has also begun to explore adapter-based foundation-model approaches in VR, in locomotion prediction rather than nuanced cognitive-state inference~\cite{ding2025multi}.

These works together provide relevant building blocks, but it remains unclear whether pretrained full-body motion representations transfer effectively to sparse VR telemetry for internal state inference. Our work addresses this question in two ways: (1) by evaluating an existing reconstruction-based route, AvatarPoser~\cite{jiang2022avatarposer} + MotionBERT~\cite{zhu2023motionbert} as a baseline, and (2) by introducing a VR-native motion adapter that operates directly on head\&hand motion without requiring full-body pose reconstruction.

\section{Method}

\subsection{Participants and Ethics}
We recruited 26 participants from our university to take part in the experiment. Data collected from 2 of them were discarded due to data quality and completeness issues. The remaining 24 participants consist of 10 females, 11 males, and 1 non-binary. They were university students or graduated students, aged 20-28 (M=23.8, SD=2.81).
Among them, 3 individuals (12.5\%) used VR devices more than once a week, 15 (62.5\%) had prior VR experience but were not frequent users, and 6 (25.0\%) had no prior VR experience.
% Recruitment and experimental procedures were approved by the university's Institutional Review Board (IRB). 
% Participants who agreed to the compensation policy received a \$\textbf{30} gift card after completing the experiment. 

All participants read and signed an IRB-approved consent form. A within-subject design was used, and all participants underwent the same experimental conditions.

\begin{figure}[tb]
    \centering
    \includegraphics[width=\linewidth]{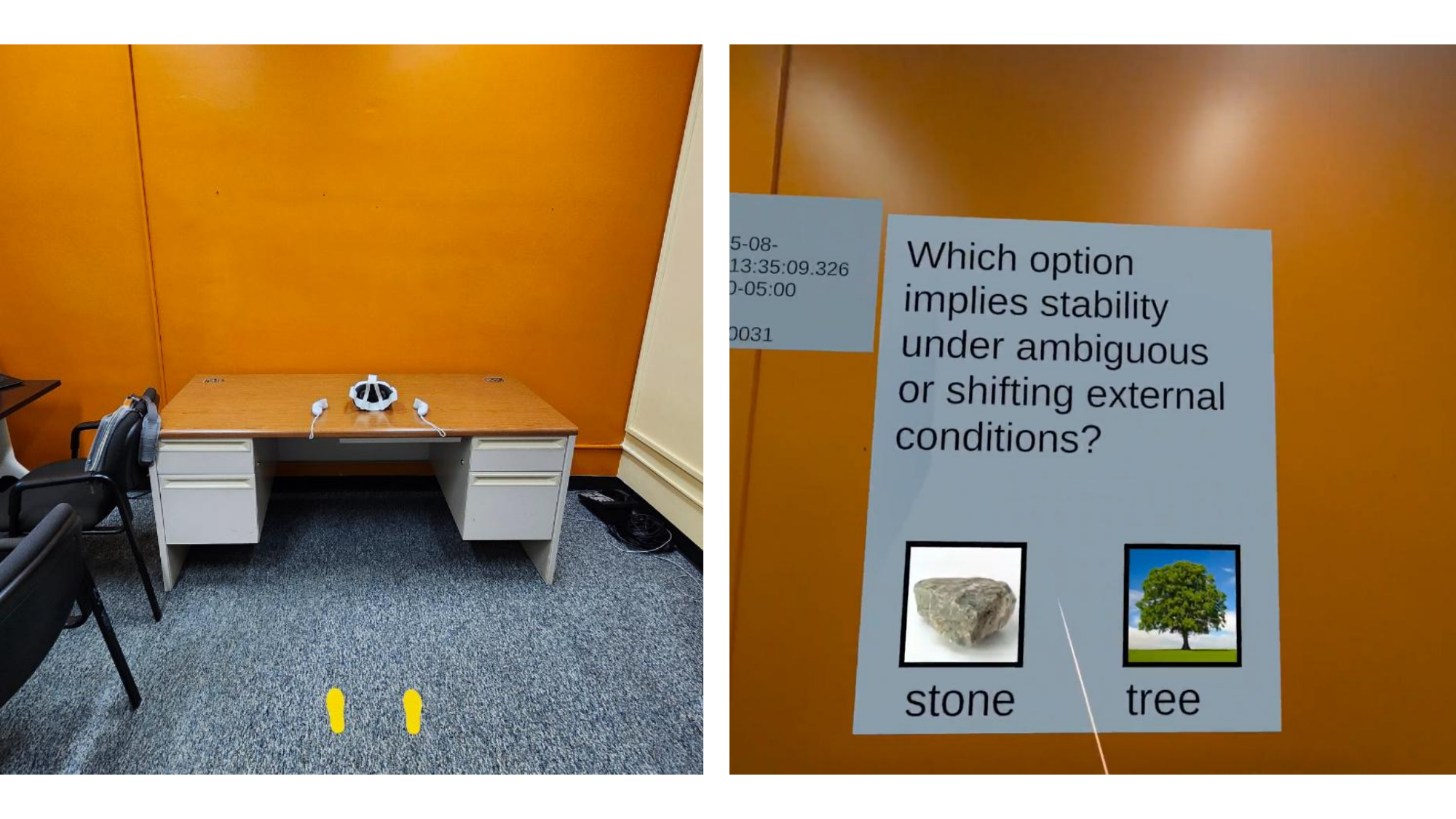}
    \caption{The experimental setup. (Left) The physical lab space where participants stood on the marked footprints. (Right) A screenshot of the in-VR view, showing an example question from the decision-making task.}
    \label{fig:expsetupp}
\end{figure}

\subsection{Apparatus}

The experiment was conducted in VR using a Meta Quest 3S HMD with two handheld controllers. The VR application was developed in Unity 2022.3.52f1. Participants stood in a fixed spot marked with stickers on the ground, wearing the HMD and holding both controllers with both hands. The VR question board was positioned directly in front of the participant.
% ; the experimenter remained at a fixed location, sitting at the left side of the participant
The experimental space setup is shown in \cref{fig:expsetupp} (left).

\subsection{Procedure}

We designed the VR task to elicit three transient decision-related cognitive states: \textit{confusion}, \textit{hesitation}, and \textit{readiness}. A short practice trial was provided before the main session. As shown in \cref{fig:expsetupp} (right), participants completed 30 two-choice conceptual questions presented in a consistent visual layout to reduce motor confounds. Each question appeared at the top of a virtual question board, with two answer options shown at the lower left and lower right. Participants read the question, considered the two options, and selected the answer they believed best matched the prompt using the controllers.

The questions were intentionally designed to evoke the target cognitive states. To do so, we included various types of questions that could induce uncertainty and delayed commitment, including unfamiliar terminology, ambiguous or misleading answer options, and subjective prompts without a clearly correct answer.
% such as questions containing potentially unfamiliar terminology, questions paired with two incorrect options, questions paired with two identical options, and subjective questions without a clearly correct answer. 
% In some cases, both options could appear plausible, making the selection between them xxx. 
This design was intended to create situations in which participants might experience confusion about the task criteria, hesitate between two options, or become ready to act once a decision had been formed, but are waiting for the desired signals to react.

To improve the reliability of the task design, we first brainstormed a diverse set of candidate questions and then refined them through two rounds of pilot experiments. After each pilot round, we reviewed pilot participants' annotations and revised the question set to better elicit confusion, hesitation, and readiness. The iteration of the question lists was intended to reduce ad hoc question design and improve the validity of our dataset.

To further reduce bias in the collected motion data, question order was randomized for each participant. In addition, the left-right placement of the two answer options was flipped with 50\% probability, so that motion patterns would not be systematically associated with a fixed answer-side layout.

% Each participant completed 30 questions in VR. Across all trials, the presentation format and interaction procedure remained the same.

\begin{figure}[tb]
    \centering
    \includegraphics[width=\linewidth]{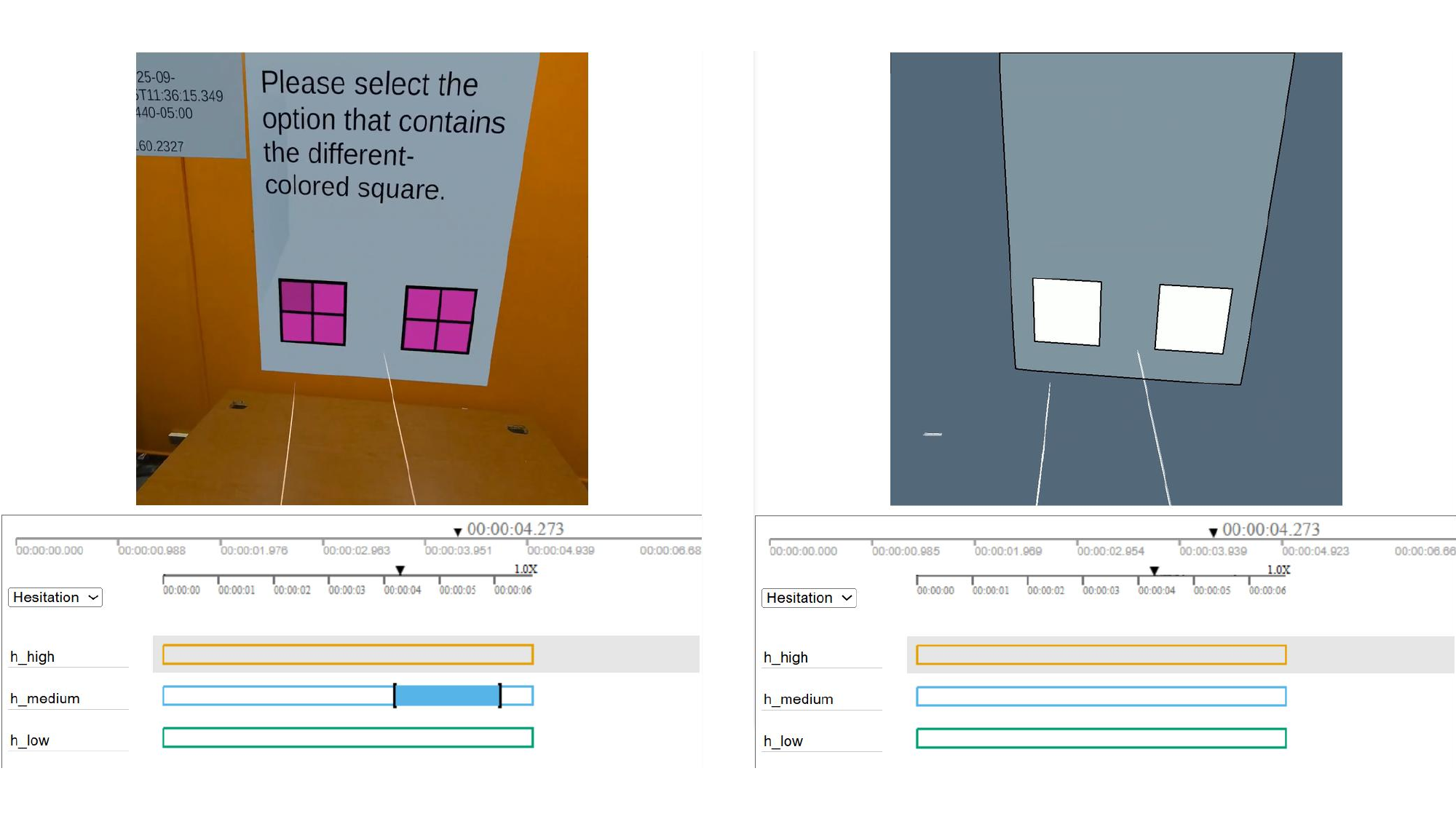}
    % \includegraphics[width=\linewidth,
    %                  trim=0 50 75 0,
    %                  clip]{figures/via.pdf}
    % \caption{The VIA annotation software. (Left) Participants doing self-annotation by watching the VR camera recording of their own. (Right) Participants doing human baseline evaluation by predicting by watching previous participant's masked videos.}
    \caption{The VIA annotation interface. (Left) Example of self-annotation, where participants reviewed their own VR session recordings. (Right) Example of human baseline evaluation, where participants inferred states from a previous participant’s masked videos, with task text and options hidden but motion preserved. In this example, the predictor failed to detect the rater's hesitation in the trial.}
    \label{fig:via}
\end{figure}

\subsection{Annotation Protocol and Label Definition}

To obtain frame-level ground truth for participants' \textit{confusion}, \textit{hesitation}, and \textit{readiness} during the VR session, we adopted a cued retrospective approach~\cite{van2005uncovering} to collect these subjective data. Prior research has shown that retrospective reporting with visual cues, such as images or videos, can support accurate recall in short-term studies~\cite{russell2009retrospective, russell2014looking, eger2007cueing}. In our study, the VR main camera recorded each participant's first-person perspective videos throughout the VR session. Immediately afterward, participants reviewed their video recordings and provided temporal annotations of the states they experienced while answering each question. 

We used the VGG Image Annotator (VIA)~\cite{dutta2016vgg} shown in \cref{fig:via}, which 
% offers video annotation capabilities, allowing 
allows participants to annotate on the timeline while reviewing their first-person videos. During annotation, participants could label four categories: \textit{Reading}, \textit{Confusion}, \textit{Hesitation}, and \textit{Readiness}. In our final modeling pipeline, \textit{Reading} was treated as a behavioral label and excluded from the cognitive-state prediction task. The final prediction target therefore consists of the three cognitive labels \textit{Confusion}, \textit{Hesitation}, and \textit{Readiness}.

During annotation, participants were given explicit instructions on how to distinguish these states. In our setting, \textit{Reading} refers to the initial period in which the participant reads the question and the two options. \textit{Confusion} refers to moments in which the participant is unsure what to do because the task criteria are unclear or ambiguous. \textit{Hesitation} refers to moments in which the participant understands the criteria, but has not yet made a decision and is still comparing the two options. \textit{Readiness} refers to moments when the participant is clear about the task and is waiting for the appropriate signal to act (e.g., waiting for the background color to change), without uncertainty or indecision. Participants were instructed that if none of these states occurred, they did not need to label anything; if multiple states occurred simultaneously, they should label each state separately.

This annotation protocol is important for the final task formulation. Our goal is not only to distinguish cognitive-state windows from non-cognitive-state windows, but also to distinguish the nuanced difference between states such as \textit{confusion} and \textit{hesitation}, which may appear similar in motion but correspond to different stages of a decision-making process.

In addition, to evaluate whether participants reach a consensus when annotating, e.g., what kinds of thinking processes should be labeled as confusion, we also collected text-based descriptions and ratings from participants. Each participant, first acting as a descriptor, wrote plain-language accounts of their decision-making process for two randomly sampled questions (see supplementary material for full instructions). These written descriptions were then shown to other participants, who acted as raters. Raters judge whether the description reflected \textit{Confusion}, \textit{Hesitation}, or \textit{Readiness} independently. 
We used these ratings to compute Cohen’s $\kappa$ as an additional measure of annotation agreement.

% the reliability of the text-based description and rating task (see \cref{text-based}), we measured agreement between each descriptor’s self-report and the two independent raters who judged whether the description reflected Confusion, Hesitation, or Readiness. Agreement was quantified using Cohen’s $\kappa$, computed per state and averaged across descriptors.

We observed substantial agreement for Confusion ($\kappa=0.69$) and Hesitation ($\kappa=0.62$), and almost perfect agreement for Readiness ($\kappa=0.95$). The overall mean across states was $\kappa=0.75$. These results indicate that participants reached an agreement when interpreting and distinguishing between confusion, hesitation, and readiness when evaluating decision-making processes, supporting the reliability of our dataset annotations.

\begin{figure}[tb]
    \centering
    \includegraphics[width=\linewidth]{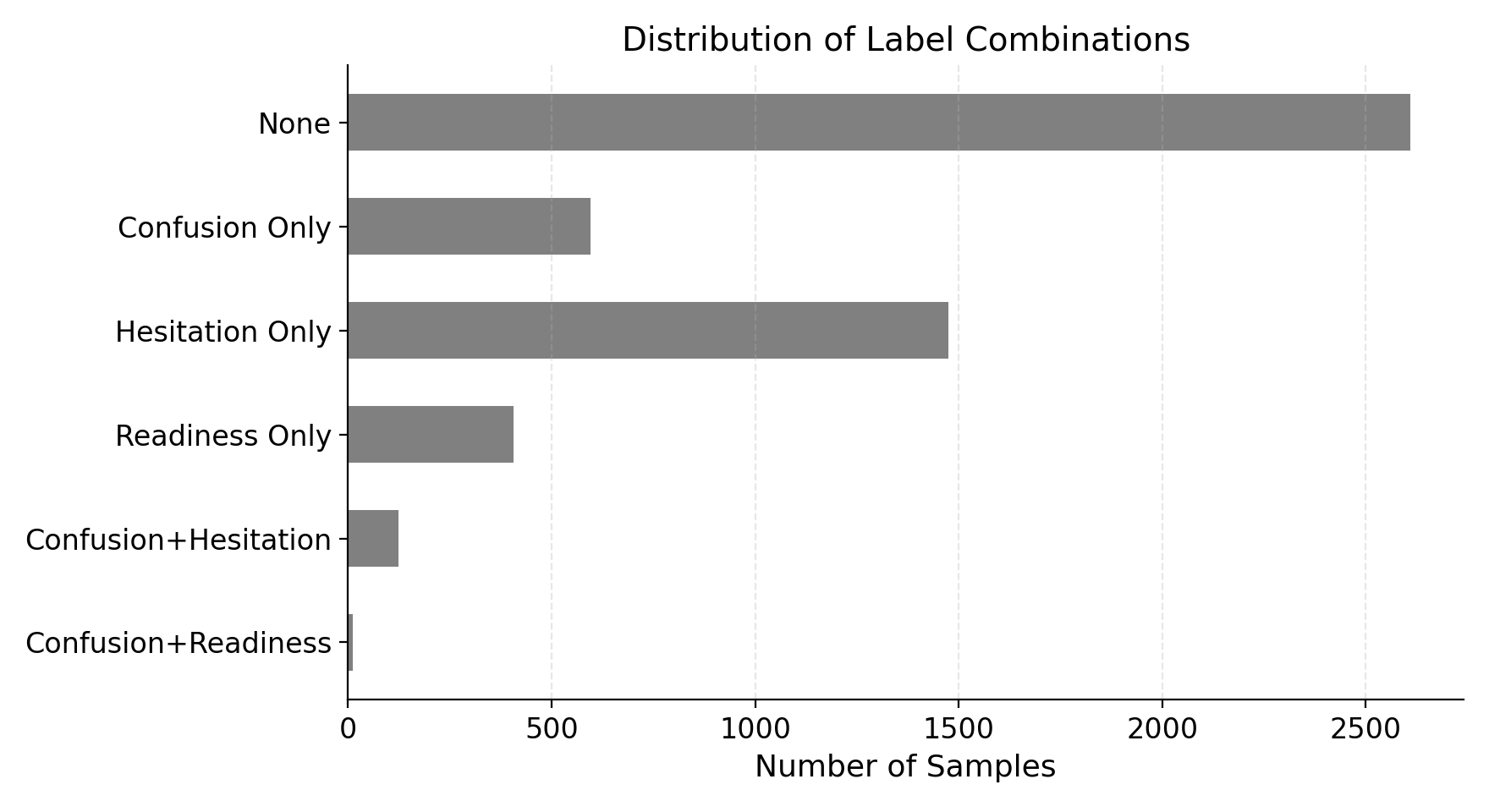}
    % \caption{Distribution of cognitive-state label combinations in the dataset. Most windows contain no labeled cognitive state or a single state, while a small number contain co-occurring states such as confusion and hesitation. This distribution illustrates both the class imbalance and the multi-label nature of the task.}
    \caption{Distribution of cognitive-state label combinations in the dataset.}
    \label{fig:dataset_distribution}
\end{figure}

\subsection{Dataset Construction}

Raw tracking data from the headset, left controller, and right controller were recorded continuously at 72\,Hz during the VR session. For each tracker, we logged timestamped 6DoF motion consisting of 3D position and 3D rotation. Positions were recorded in meters, and rotations were stored as Euler angles in degrees using Unity’s default coordinate system.

The custom-built Unity application additionally recorded trial timestamps and a continuous first-person VR camera video. After the session, this video was segmented into 30 trial-specific clips using the Unity timestamps. These clips were then uploaded to VIA~\cite{dutta2016vgg} for retrospective self-annotation. 
% To ensure data completeness, we required each valid session to contain exactly 30 trials, all three trackers' raw motion data, one VR recording, one self-annotation file, and one human-baseline annotation file (except for the first participant, for whom no previous-participant annotation exists). 
% More details regarding the human baseline are described in \cref{human_baseline}.

Before constructing the dataset, the three tracker streams were synchronized to a shared 72\,Hz timeline. Euler angles were first unwrapped along each axis to remove $360^\circ$ discontinuities. The head, left-hand, and right-hand streams were then linearly interpolated onto a common 72\,Hz grid. Short missing segments of at most two frames were forward-filled, while longer missing segments were discarded to preserve signal quality.

Temporal cognitive-state annotations were rasterized onto the same 72\,Hz timeline to obtain frame-level labels. 
% Although participants annotated four categories---\textit{Reading}, \textit{Confusion}, \textit{Hesitation}, and \textit{Readiness}---the final prediction task focuses only on the three cognitive-state labels \textit{Confusion}, \textit{Hesitation}, and \textit{Readiness}. 
We construct our dataset by converting the continuous timeline into windows of width 2.0\,s with a stride of 0.5\,s, yielding windows of 144 frames each. Frame-level labels were then converted into window-level multilabel targets using a simple thresholding: a window was labeled positive for each cognitive state if at least 20\% of its frames were labeled as positive.

This process yields a three-label multilabel binary classification task, where each window is assigned a target vector $y \in \{0,1\}^3$ corresponding to \textit{Confusion}, \textit{Hesitation}, and \textit{Readiness}. Because these labels are predicted jointly, models must distinguish not only cognitive-state windows from non-cognitive-state windows, but also subtle differences among related states, especially \textit{Confusion} and \textit{Hesitation}. Since naive sliding-window extraction overproduces long negative stretches and repeated windows from long positive events, we further apply event-capped sampling. For each contiguous positive event of a given label, we retain at most six stride-aligned positive windows. We then sample pure-negative windows so that the number of all-zero windows matches the number of windows containing at least one positive label.

\cref{fig:dataset_distribution} visualizes the distribution of label combinations across the resulting 2\,s windows. Most windows contain either no labeled cognitive state or a single state, while a smaller portion contain co-occurring states such as \textit{Confusion} and \textit{Hesitation}. The figure highlights both the class imbalance of the dataset and the multi-label nature of the task. In particular, \textit{Hesitation} appears more frequently than the other cognitive states.

\subsection{Feature Extraction}

From the synchronized tracker streams, we constructed a 69-dimensional motion descriptor for each frame. For each of the three tracked objects (head, left hand, and right hand), we extracted raw position, raw rotation in Euler angles, linear velocity, angular velocity, linear acceleration, and angular acceleration, yielding 18 dimensions per tracker and 54 dimensions in total. We further computed 15 body-centered relational features: for each hand, we included its 3D position and 3D rotation relative to the head together with its Euclidean distance to the head, yielding 7 features per hand, and we added one symmetry feature defined as the difference between the left-hand and right-hand distances to the head. This resulted in a final 69-dimensional feature vector for each frame.

This 69D per-frame feature serves as the shared processed motion stream for the classical and temporal baselines. It provides a compact but information-rich summary of sparse VR motion, capturing both instantaneous pose and local kinematics.

The MotionBERT-based models do not consume this representation directly, since MotionBERT~\cite{zhu2023motionbert} expects pose-like joint sequences rather than engineered tracker features. Instead, both MotionBERT-based models start from the same synchronized sparse VR motion but use different adaptation pipelines.

For AvatarPoser--MotionBERT, we construct an AvatarPoser-compatible 54-dimensional representation from the sparse tracker signals. Starting from head, left-hand, and right-hand position and rotation signals, we resample the motion from 72\,Hz to 60\,Hz, convert Euler rotations to 6D rotation representations, compute inter-frame rotation velocity and position differences, and concatenate the resulting signals into the format required by AvatarPoser~\cite{jiang2022avatarposer}. This representation is used because AvatarPoser is designed as a sparse-to-full-body reconstruction model and requires tracker-based motion features aligned with its input format.

For the proposed VRMotionBERT, we construct a VR-native token representation directly from sparse tracking data. In the reported best configuration, each frame is represented by three tracker tokens corresponding to the head, left hand, and right hand. Each token contains a head-centered position, a 6D absolute rotation converted from Euler angles, linear velocity, and angular velocity, yielding a 15-dimensional token per tracker. The resulting input window has shape $(144, 3, 15)$. We use this representation because it preserves the native sparse-VR structure while allowing the proposed adapter to learn a direct mapping from tracker motion to MotionBERT-compatible pseudo joints without explicit full-body pose reconstruction.

\subsection{Human Baseline}
\label{human_baseline}

We also evaluate a human-observer baseline to measure how well people can infer cognitive states from VR motion alone and to provide a reference point for model performance. As shown in \cref{fig:via} (right), each participant was asked to infer the cognitive states of another participant rather than their own by watching that participant's recorded VR session.

To ensure a fair comparison with the models, we masked all question and answer content in the videos used for the human-baseline evaluation. As a result, observers could infer \textit{confusion}, \textit{hesitation}, and \textit{readiness} only from the other participant's head\&hand motion, without access to the semantic content of the task.

Notably, this human baseline remains cross-person even under the same-user evaluation setting reported in \cref{tab:seen}. 
Human performance was evaluated using the same 72Hz timeline, the same thresholding, and the same test windows used for model comparisons.
This procedure ensures a fair comparison between human judgments and model predictions.

\subsection{Model Families}

We evaluate five models spanning classical machine learning, temporal neural networks, and pretrained motion foundation models.

\paragraph{SVM.}
As a classical baseline, we use a linear multi-output support vector machine. Each 2s window is flattened and classified with three simultaneous binary outputs. This baseline tests how far static linear decision boundaries can go when given the full processed window.

\paragraph{DTCN.}
We use the deep temporal convolutional network (DTCN)~\cite{koh2021deep} as a convolutional sequence baseline. DTCN operates directly on the $(144,69)$ sequence and represents deep temporal modeling of the motion sequence.

\paragraph{MLSTM-FCN.}
We use the multivariate LSTM-FCNs (MLSTM-FCN)~\cite{karim2019multivariate} as a hybrid recurrent-convolutional temporal baseline. Similar to DTCN, it operates directly on the shared $(144,69)$ sequence representation, but combines recurrent modeling with convolutional feature extraction.

\paragraph{AP--MotionBERT.}
We implemented AP--MotionBERT as a reconstruction-based baseline that adapts sparse VR motion to MotionBERT~\cite{zhu2023motionbert} through AvatarPoser~\cite{jiang2022avatarposer}. To match AvatarPoser’s input requirements, we extract tracker-specific position and rotation signals, resample them from 72\,Hz to 60\,Hz, convert Euler rotations to 6D rotation representations, and compute relative rotation velocity and position velocity, yielding an AvatarPoser-compatible 54-dimensional sequence. AvatarPoser then reconstructs a dense full-body pose sequence from sparse head-and-hand motion. The reconstructed pose is then mapped into a 17-joint body-keypoint representation compatible with MotionBERT and passed through a pretrained MotionBERT backbone used as a feature extractor. The high-dimensional motion representation by MotionBERT is then fed into a trainable temporal classifier head for the multi-label binary classification task. 

In the reported AP--MotionBERT pipeline, both AvatarPoser and MotionBERT are initialized with publicly available pretrained weights and kept frozen. Only the temporal classifier head is optimized to map MotionBERT’s output representations to the three target labels. This baseline therefore represents a reconstruction-based route for transferring sparse VR motion into a full-body motion foundation model. This reflects the typical usage of AvatarPoser as a pretrained sparse-to-full-body reconstruction module and avoids overfitting given the limited size of our VR dataset.

\paragraph{VRMotionBERT (VR-native MotionBERT).}
Our proposed method, VRMotionBERT, removes the explicit full-body reconstruction stage and instead learns a direct VR-native interface to MotionBERT. Starting from the same 69-dimensional processed motion stream, we construct three tracker tokens per frame, corresponding to the head, left hand, and right hand. In the reported best configuration, each token contains: 3D head-centered position, 6D absolute rotation converted from Euler angles, 3D linear velocity, and 3D angular velocity, yielding a 15-dimensional token per tracker and an input window of shape
$(144, 3, 15)$.

The proposed VR motion adapter utilizes a lightweight MLP-based architecture that maps the sparse tracker tokens into MotionBERT-compatible pseudo-joint features. A learned coordinate projection then converts these pseudo-joint features into a 17-joint pseudo-skeleton in 3D, which is passed to MotionBERT. Unlike AP--MotionBERT, VRMotionBERT is trained jointly through the adapter, the pseudo-joint projection, the upper MotionBERT layers, and the temporal classifier head. In the reported best configuration, the lower MotionBERT layers remain frozen while the top three block pairs are unfrozen and fine-tuned.

% A key difference between AP-MotionBERT and VRMotionBERT is therefore the adaptation strategy. AP-MotionBERT first reconstructs a full-body pose using an external motion reconstructor and then applies MotionBERT features, while VRMotionBERT directly learns a representation compatible with MotionBERT without explicit full-body pose reconstruction. \cref{fig:comparing} illustrates this architectural difference.

\begin{figure*}[tb]
    \centering
    \includegraphics[width=\linewidth]{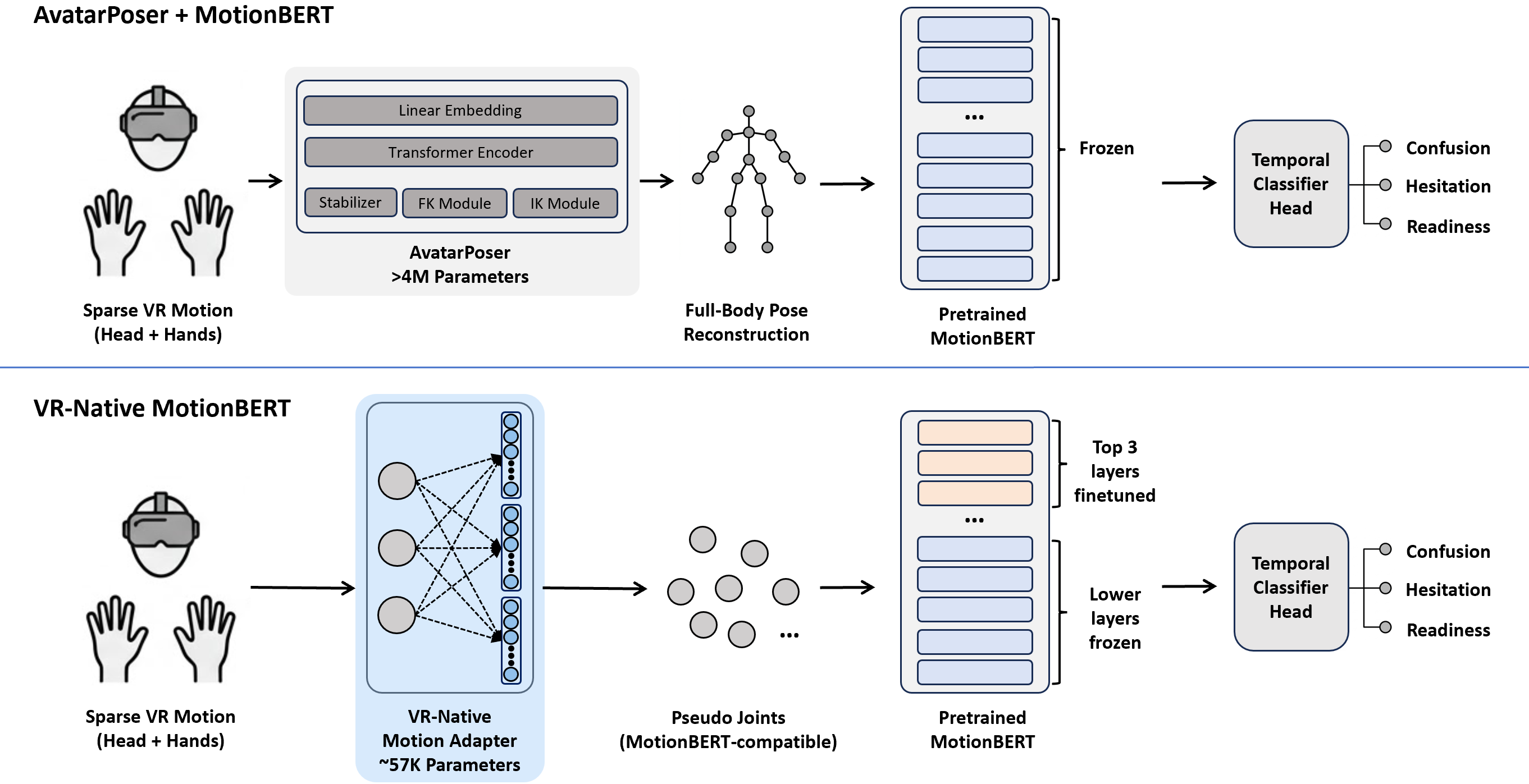}
    \caption{Comparison between model architecture and parameter count of AP--MotionBERT and VRMotionBERT. }
    \label{fig:comparing}
\end{figure*}

The two approaches differ primarily in how they represent motion before feeding it into MotionBERT. As shown in \cref{fig:comparing}, the AvatarPoser + MotionBERT pipeline first reconstructs a full-body pose from sparse VR inputs using a large pose estimation model, introducing an additional intermediate step with over 4M parameters. In contrast, VR-native MotionBERT directly learns a lightweight motion adapter (~57K parameters) that maps sparse head and hand signals into MotionBERT-compatible pseudo joints. This design removes the need for full-body reconstruction, resulting in a more compact, task-aligned representation while significantly reducing model complexity and potential sources of noise.

\subsection{Training Setup}

For all neural models, we use weighted binary cross-entropy loss over the three target cognitive states. Models are trained with the Adam optimizer~\cite{kingma2014adam}, using a learning rate of $3 \times 10^{-4}$ and weight decay of $10^{-4}$. Gradient norms are clipped to 1.0, and automatic mixed precision is enabled during training. Each run trains for 100 epochs. All experiments use a fixed random seed of 42.

Model-specific settings follow the finalized repository configuration. DTCN uses four residual dilated convolution blocks with 128 hidden channels, kernel size 4, and dropout 0.2. MLSTM-FCN uses a one-layer LSTM with a hidden size of 128, together with a convolutional branch. For both AP--MotionBERT and VRMotionBERT, we use the MotionBERT-Lite backbone followed by a trainable temporal classifier head that maps MotionBERT sequence representations to three output logits.

\subsection{Evaluation Protocol and Metrics}
\label{subsec:evaluation_protocol}
We evaluate all models under two protocols.

\textbf{Future-in-time prediction for the same users.}
For every participant, the first 70\% of the timestamps are used to construct the training set, while the remaining 30\% is used for testing. It is guaranteed that there are no overlapping windows in the training and test sets. This protocol evaluates whether models can predict future cognitive states from VR motion for users already observed during training.

\textbf{Generalization to previously unseen users.}
We perform 6-fold cross-validation over participants, holding out four participants in each fold for testing and using the remaining participants for training. This protocol evaluates models' generalization ability to previously unseen users, an essential property for practical deployment in real-world VR applications.

In both protocols, the task is formulated as multi-label binary classification with three simultaneous outputs: \textit{confusion}, \textit{hesitation}, and \textit{readiness}. We report overall and per-label Accuracy, Precision, Recall, and F1. For the unseen-user protocol, we report mean $\pm$ standard deviation across folds to reflect both predictive performance and cross-fold stability. Human-baseline performance is computed on the same test windows, enabling a fair comparison between human judgments and model predictions.

\subsection{Implementation and Reproducibility}

Experiments were conducted on Ubuntu 24.04.3 LTS. Model training and evaluation were implemented in Python 3.8.20 and PyTorch 2.4.1~\cite{paszke2019pytorch} on a workstation equipped with an NVIDIA RTX 3080 GPU. Our dataset and modeling framework will be made publicly available to support future research.

\begin{table*}[t]
\centering
% \caption{Setting 1 - Pooled features, binary classification, 70/30 split. Input: $[B,105]$, Output: $[B,8]$. The highest value is highlighted in \textbf{bold}.}
\caption{Setting 1 - Future-in-time prediction for the same users, 70/30 split. \textit{TBD: bold best model result, or best among human and models.}
\textbf{Bold} indicates the best model performance.
$^\star$ denotes results surpassing the human baseline.
}

\label{tab:seen}
\setlength{\tabcolsep}{4pt}
\renewcommand{\arraystretch}{1.2}
\resizebox{1.0\linewidth}{!}{
% \begin{tabular}{l cccc cccc cccc cccc cccc}
\begin{tabular}{l llll llll llll llll}

\toprule
\multirow{2}{*}{\textbf{Model}} &
\multicolumn{4}{c}{\textbf{Overall}} &
% \multicolumn{5}{c}{\textbf{Reading}} &
\multicolumn{4}{c}{\textbf{Confusion}} &
\multicolumn{4}{c}{\textbf{Hesitation}} &
\multicolumn{4}{c}{\textbf{Readiness}} \\
 % & Acc. & Prec. & Rec. & F1  
 % & Acc. & Prec. & Rec. & F1 
 % & Acc. & Prec. & Rec. & F1
 % & Acc. & Prec. & Rec. & F1  \\
& \multicolumn{1}{c}{Acc.} 
& \multicolumn{1}{c}{Prec.} 
& \multicolumn{1}{c}{Rec.} 
& \multicolumn{1}{c}{F1}  

& \multicolumn{1}{c}{Acc.} 
& \multicolumn{1}{c}{Prec.} 
& \multicolumn{1}{c}{Rec.} 
& \multicolumn{1}{c}{F1} 

& \multicolumn{1}{c}{Acc.} 
& \multicolumn{1}{c}{Prec.} 
& \multicolumn{1}{c}{Rec.} 
& \multicolumn{1}{c}{F1}

& \multicolumn{1}{c}{Acc.} 
& \multicolumn{1}{c}{Prec.} 
& \multicolumn{1}{c}{Rec.} 
& \multicolumn{1}{c}{F1} \\
\midrule

% Human & \textbf{0.793} & \textbf{0.458} & 0.455 & \textbf{0.431} & \textbf{0.344}
%       & \textbf{0.850} & \textbf{0.804} & \textbf{0.903} & \textbf{0.839} & \textbf{0.728}
%       & 0.747 & 0.282 & 0.242 & 0.243 & \textbf{0.174}
%       & 0.658 & 0.621 & 0.556 & 0.543 & 0.409
%       & \textbf{0.916} & 0.124 & 0.119 & 0.099 & 0.066 \\
\textit{Human}&0.810&0.537&0.574&0.555&0.802&0.244&0.262&	0.253&0.733&0.635&0.734&0.681&0.894&0.480&0.353	&0.407\\
\midrule

SVM & 0.637 & 0.197 & 0.334 & 0.248 & 0.628 & 0.095 & 0.245 & 0.131 & 0.573 & 0.362 & 0.417 & 0.388 & 0.708 & 0.069 & 0.197 & 0.102             \\

DTCN & 0.687 & 0.300 & \textbf{0.558} & 0.390 & 0.716 & 0.150 & \textbf{0.252} & 0.188 & 0.517 & 0.379 & \textbf{0.772}$^\star$ & 0.509 & 0.828 & 0.144 & 0.212 & 0.171 \\

MLSTM-FCN & 0.790 & 0.418&0.436&0.427&0.833$^\star$&0.248$^\star$&0.136&0.175&0.681&0.506&0.608&0.552&0.855&\textbf{0.199}&\textbf{0.241}&\textbf{0.218}\\

AP--MotionBERT & 0.765 & 0.358 & 0.385 & 0.371 & \textbf{0.863}$^\star$ & 0.222 & 0.019 & 0.034 & 0.533 & 0.369 & 0.625 & 0.464 & \textbf{0.900}$^\star$ & 0.118 & 0.029 & 0.047 \\

\midrule
VRMotionBERT & \textbf{0.820}$^\star$ & \textbf{0.500}  & 0.449 & \textbf{0.473} &  0.847$^\star$ & \textbf{0.347}$^\star$ & 0.192 & \textbf{0.247} & \textbf{0.725} & \textbf{0.566} & 0.651 & \textbf{0.605} & 0.889 & 0.156 & 0.073 & 0.100                  \\

\bottomrule
\end{tabular}}
\end{table*}

\begin{table*}[t]
\centering
\caption{Setting 2 - Generalization to previously unseen users, 6-fold cross-validation. \textit{TBD: bold best model result, or best among human and models.}
\textbf{Bold} indicates the best model performance.
$^\star$ denotes results surpassing the human baseline.}
\label{tab:unseen_6fold_reformatted}

\setlength{\tabcolsep}{4pt}
\renewcommand{\arraystretch}{1.15}

%==================== Top half: Overall + Confusion ====================%
\resizebox{1\linewidth}{!}{
\begin{tabular}{l cccc cccc}
\toprule
\multirow{2}{*}{\textbf{Model}} &
\multicolumn{4}{c}{\textbf{Overall}} &
\multicolumn{4}{c}{\textbf{Confusion}} \\
& \textbf{Acc.} & \textbf{Prec.} & \textbf{Rec.} & \textbf{F1}
& \textbf{Acc.} & \textbf{Prec.} & \textbf{Rec.} & \textbf{F1} \\
\midrule

\textit{Human}
& 0.802$\pm$0.061 & 0.481$\pm$0.129 & 0.486$\pm$0.160 & 0.481$\pm$0.139
& 0.798$\pm$0.087 & 0.355$\pm$0.213 & 0.370$\pm$0.183 & 0.356$\pm$0.195 \\
\midrule

SVM
& 0.639$\pm$0.020 & 0.201$\pm$0.022 & 0.357$\pm$0.035 & 0.257$\pm$0.026
& 0.658$\pm$0.040 & 0.122$\pm$0.028 & 0.225$\pm$0.037 & 0.156$\pm$0.026 \\
DTCN
& 0.685$\pm$0.053 & 0.243$\pm$0.041 & 0.361$\pm$0.096 & 0.283$\pm$0.041
& 0.667$\pm$0.124 & 0.178$\pm$0.080 & \textbf{0.309$\pm$0.222} & 0.170$\pm$0.081 \\
MLSTM-FCN
& 0.746$\pm$0.030 & 0.350$\pm$0.023 & \textbf{0.509$\pm$0.100}$^\star$ & 0.410$\pm$0.027
& 0.777$\pm$0.031 & 0.244$\pm$0.062 & 0.287$\pm$0.153 & \textbf{0.249$\pm$0.080} \\
AP--MotionBERT
& 0.761$\pm$0.011 & 0.339$\pm$0.028 & 0.389$\pm$0.058 & 0.361$\pm$0.036
& 0.809$\pm$0.028$^\star$ & 0.261$\pm$0.076 & 0.155$\pm$0.055 & 0.178$\pm$0.041 \\
\midrule
VRMotionBERT
& \textbf{0.820$\pm$0.023}$^\star$ & \textbf{0.492$\pm$0.068}$^\star$ & 0.460$\pm$0.071 & \textbf{0.471$\pm$0.053}
& \textbf{0.842$\pm$0.015}$^\star$ & \textbf{0.399$\pm$0.119}$^\star$ & 0.183$\pm$0.062 & 0.239$\pm$0.054 \\
\bottomrule
\end{tabular}
}

\vspace{0.2em}

%==================== Bottom half: Hesitation + Readiness ====================%
\resizebox{1\linewidth}{!}{
\begin{tabular}{l cccc cccc}
\toprule
\multirow{2}{*}{\textbf{Model}} &
\multicolumn{4}{c}{\textbf{Hesitation}} &
\multicolumn{4}{c}{\textbf{Readiness}} \\
& \textbf{Acc.} & \textbf{Prec.} & \textbf{Rec.} & \textbf{F1}
& \textbf{Acc.} & \textbf{Prec.} & \textbf{Rec.} & \textbf{F1} \\
\midrule

\textit{Human}
& 0.720$\pm$0.075 & 0.586$\pm$0.106 & 0.594$\pm$0.199 & 0.580$\pm$0.138
& 0.888$\pm$0.051 & 0.237$\pm$0.225 & 0.212$\pm$0.208 & 0.205$\pm$0.185 \\
\midrule

SVM
& 0.528$\pm$0.028 & 0.312$\pm$0.034 & 0.462$\pm$0.083 & 0.367$\pm$0.024
& 0.731$\pm$0.040 & 0.079$\pm$0.028 & \textbf{0.214$\pm$0.077}$^\star$ & 0.114$\pm$0.039 \\
DTCN
& 0.559$\pm$0.039 & 0.324$\pm$0.051 & 0.428$\pm$0.113 & 0.362$\pm$0.062
& 0.829$\pm$0.093 & 0.102$\pm$0.072 & 0.186$\pm$0.173 & 0.125$\pm$0.096 \\
MLSTM-FCN
& 0.559$\pm$0.075 & 0.382$\pm$0.045 & 0.706$\pm$0.146 & 0.488$\pm$0.036
& \textbf{0.903$\pm$0.019}$^\star$ & \textbf{0.337$\pm$0.174}$^\star$ & 0.151$\pm$0.104 & 0.193$\pm$0.105 \\
AP--MotionBERT
& 0.583$\pm$0.020 & 0.371$\pm$0.039 & 0.563$\pm$0.089 & 0.445$\pm$0.048
& 0.890$\pm$0.027$^\star$ & 0.195$\pm$0.103 & 0.137$\pm$0.094 & 0.153$\pm$0.090 \\
\midrule
VRMotionBERT
& \textbf{0.721$\pm$0.044}$^\star$ & \textbf{0.536$\pm$0.072} & \textbf{0.652$\pm$0.083}$^\star$ & \textbf{0.583$\pm$0.059}$^\star$
& 0.897$\pm$0.033$^\star$ & 0.246$\pm$0.197$^\star$ & 0.206$\pm$0.164 & \textbf{0.205$\pm$0.154}$^\star$ \\
\bottomrule
\end{tabular}
}
\end{table*}

\begin{figure*}[tb]
    \centering
    \includegraphics[width=\linewidth]{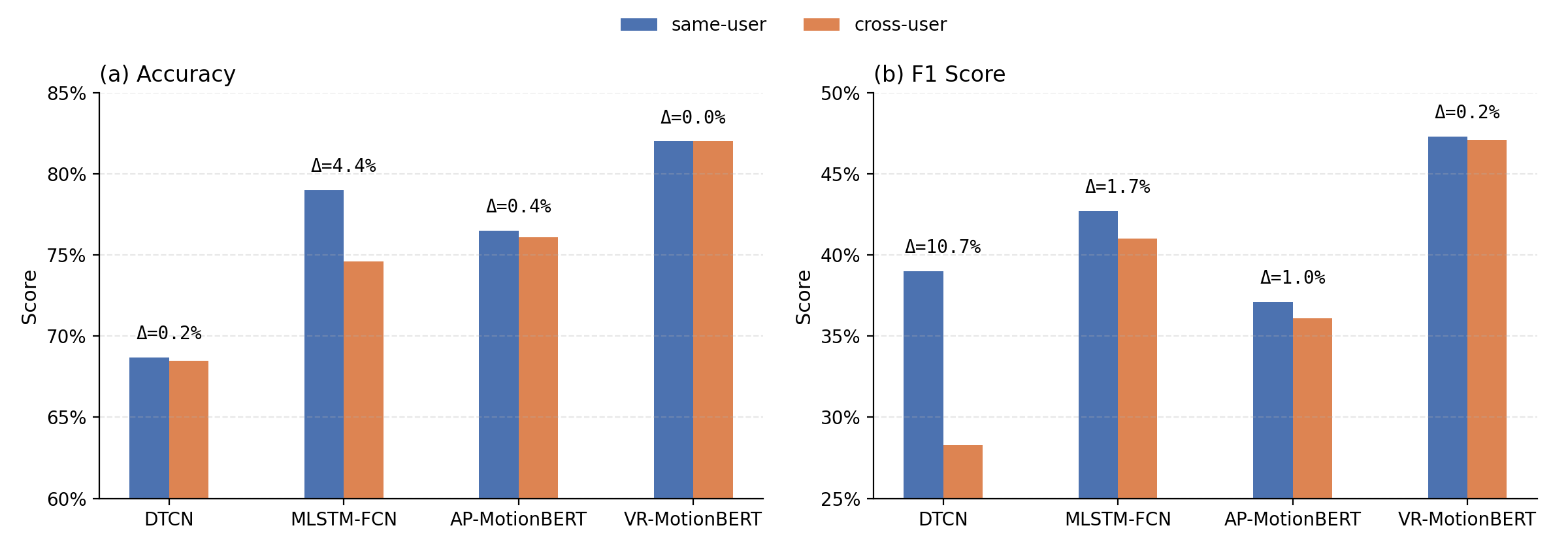}
    \caption{Comparison of model (a) accuracy and (b) F1-score under same-user and cross-user evaluation protocols. $\Delta$ denotes the performance drop for each model when generalizing to new (unseen) users.}
    \label{fig:seen_unseen_performance_drop}
\end{figure*}

% \begin{table}[t]
% \centering
% \caption{Positive-label ratio (\%) for each cognitive state in the same-user setting. The training and test sets are partitioned by a 70/30 temporal split for each participant to simulate real-world adaptation to known users' future behavior.}
% \label{tab:label_distribution}
% \setlength{\tabcolsep}{10pt}
% \renewcommand{\arraystretch}{1.15}
% \begin{tabular}{lcc}
% \toprule
% \textbf{Label} & \textbf{Train} & \textbf{Test} \\
% \midrule
% Confusion  & 14.4\% & 13.1\% \\
% Hesitation & 29.8\% & 32.4\% \\
% Readiness  & 7.8\%  & 8.4\% \\
% \bottomrule
% \end{tabular}
% \end{table}

% \textcolor{blue}{Kaiang: Editing Results finished (Mar 16)}

\section{Results}

We evaluate whether subtle cognitive states can be inferred from sparse VR motion and whether motion foundation models provide advantages over classical approaches. Specifically, we investigate three questions: 
(1) how accurately different models detect cognitive states from VR motion, 
(2) how well different models generalize to previously unseen users, and 
(3) how model performance compares with human observers. 
We report results using the two protocols described in \cref{subsec:evaluation_protocol}.

\subsection{Future-in-Time Prediction for the Same Users}

We first evaluate the models' ability to predict future cognitive states for users already observed during training. In this temporal split protocol, each participant's data is partitioned chronologically, with the earlier 70\% utilized for training and the later 30\% for testing. This ensures that no overlapping windows exist between the two sets. 
The results for this protocol are detailed in \cref{tab:seen}. Our analysis revealed the following:

Classical and convolutional models, including SVM~\cite{hearst1998support} and DTCN~\cite{koh2021deep}, achieve 60+\% overall accuracy and show a baseline capability to detect cognitive states from sparse motion. SVM in particular achieves only 19.7\% precision, indicating its high false positives. Although DTCN achieves relatively low overall accuracy, it gets the best recall of 55.8\%. This could be because of over-predicting positive samples. MLSTM-FCN~\cite{karim2019multivariate} achieves notable performance in this setting, with 79.0\% overall accuracy and the best precision, recall, and F1-score in the \textit{Readiness} category.

% Classical machine learning baselines, including SVM~\cite{hearst1998support}, DTCN~\cite{koh2021deep}, and MLSTM-FCN~\cite{karim2019multivariate}, demonstrate a baseline capability to detect cognitive states from sparse motion. This confirms that standard head and hand movement trajectories contain measurable behavioral cues linked to decision-making processes. However, these models exhibit limited performance, particularly when tasked with identifying more subtle or less frequent states.

For Motion Foundation Models, both MotionBERT-based architectures achieve relatively stronger performance than SVM and DTCN across nearly all reported metrics. This performance gap suggests that pretrained motion representations are highly effective at capturing the complex temporal structures of human movement necessary for cognitive-state inference.

Among all evaluated models, the proposed VRMotionBERT achieved the strongest overall performance, reaching an accuracy of 0.820 and an F1-score of 0.473. Notably, VRMotionBERT surpassed the human-observer baseline in overall accuracy (0.810) and accurately identified Confusion and Hesitation with higher precision than the human rater. Besides, VRMotionBERT outperformed the reconstruction-based AP-MotionBERT in both accuracy and F1-score. This indicates that the proposed VR motion adapter utilizes motion foundation models in a more effective pathway than requiring explicit full-body pose reconstruction through an external module.

These findings suggest that while motion signals are inherently subtle, foundational motion models can successfully leverage large-scale pretraining to decode internal user states in VR even with limited task-specific data.

\subsection{Generalization to Previously Unseen Users}

We next evaluate whether models generalize to users who were not observed during training. This setting is particularly important for real-world VR applications, where systems must often generalize to new individuals without the opportunity for prior personalization.

\cref{tab:unseen_6fold_reformatted} presents the results under the 6-fold cross-validation unseen-user protocol. As expected, performance decreases for all models compared to the seen-user setting, reflecting the inherent challenge of cross-person generalization in behavioral inference tasks.

The relative ranking of models changes noticeably in this challenging setting. Classical baselines exhibit substantial performance degradation when evaluated on new users, indicating a limited ability to capture user-invariant motion patterns associated with subtle cognitive states. For example, the overall F1-scores of DTCN drop from $0.390$ to $0.283$.

% Motion Foundation Model Performance. 
In contrast, motion foundation models maintain significantly stronger performance. VRMotionBERT achieves the best overall results with an accuracy of $0.820 \pm 0.023$ and an F1-score of $0.471 \pm 0.053$, maintaining comparable performance against the human baseline. This suggests that pretrained motion representations, combined with a VR-native adaptation pipeline, help the model capture behavioral features that generalize more robustly across diverse individuals. State-wise, VRMotionBERT also achieves the highest accuracy and precision in \textit{Confusion} and \textit{Hesitation}, and the highest F1-score in \textit{Readiness}, outperforming human raters in these metrics.

This suggests that pretrained motion representations, combined with a VR-native adaptation pipeline, help capture behavioral features that generalize more robustly across individuals.

\subsection{Comparison with Human Observers}

We further compare model performance with a human-observer baseline. In this evaluation, participants inferred the cognitive states of another participant solely by observing their head and hand movements, without access to the question content or answer options.

Human predictions were processed using the same windowing and labeling procedure as model outputs to ensure a fair comparison.

The results show that motion foundation models approach human-level performance in detecting cognitive states from motion alone. In particular, VRMotionBERT achieves performance comparable to, and in some cases higher than, the human baseline for several metrics. This finding suggests that subtle behavioral cues contained in sparse VR motion are sufficiently informative for cognitive state inference.

Notably, human observers showed higher variance compared to models. For example, the human baseline has a std of $0.061$ and $0.139$ in overall accuracy and F1-score, respectively, which are higher than those of all evaluated models. Additionally, though sometimes surpassed by models in accuracy, the human baseline achieved the highest F1-score in all three cognitive states. 
This suggests that human observers might be better at capturing high-level semantic patterns in motion, particularly when distinguishing nuanced differences between states such as \textit{confusion} and \textit{hesitation}.
% This suggests that human are better at the high-level understanding of motion, and identifying the nuanced differences between states like \textit{confusion} and \textit{hesitation}.

\subsection{Performance Drop from Seen to Unseen Users}

To further analyze model robustness, we examine the change in performance between the same-user and unseen-user protocols. \cref{fig:seen_unseen_performance_drop} visualizes the performance drop for each model on overall accuracy (left) and F1-score (right) separately.

Baseline models show a pronounced decrease in both accuracy and F1 when evaluated on unseen users, indicating that these models tend to rely on user-specific motion patterns. In contrast, motion foundation models exhibit a much smaller performance drop, with no drop in accuracy and $0.2\%$ in F1-score, suggesting pretrained motion foundation models' superiority in generalization.
% to unseen users.

VRMotionBERT demonstrates the most stable behavior across the two protocols. This result supports the hypothesis that directly adapting sparse VR tracking signals into a motion foundation model allows the system to leverage pretrained motion priors while remaining robust to user-specific variation.

\section{Discussion}
\label{sec:discussion}

\subsection{Main Findings}

Overall, the results support three main observations. First, subtle cognitive states such as \textit{confusion}, \textit{hesitation}, and \textit{readiness} can be inferred from sparse VR head\&hand motion alone. Second, pretrained motion foundation models consistently outperform classical machine learning and temporal models on this task. Third, the proposed VR-native adaptation approach improves generalization to previously unseen users while avoiding the need for explicit full-body reconstruction, even with a small dataset.

These observations are supported by consistent trends across both evaluation protocols. Under the within-participant setting, classical and convolutional models achieve relatively limited performance, while temporal and motion foundation models approach human-level performance. Under cross-user evaluation, classical and temporal models exhibit a noticeable performance drop, highlighting the challenge of subject-independent inference. In contrast, MotionBERT-based approaches, particularly the proposed VRMotionBERT, maintain robust performance comparable to the human baseline, indicating stronger generalization across users.

Together, these findings provide evidence that standard VR motion telemetry encodes meaningful signals of nuanced cognitive states. They further suggest that large-scale motion pretraining offers a promising pathway for XR inference tasks.
% , enabling models to capture user-invariant patterns from sparse tracking signals.

% \subsection{Main Findings}

% Our results provide the first systematic evidence that standard VR motion telemetry encodes meaningful signals of nuanced cognitive states. When models were evaluated under the within-participant protocol, classical and convolutional models achieved relatively low performance, while temporal and motion foundation models approached human-level performance. When generalizing to previously unseen users, the observed performance drop for classical, convolutional, and temporal models underscores the persistent challenge of subject-independent inference. However, MotionBert-based methods, especially the proposed VRMotionBRET, preserve robust performance comparable to the human baseline even in the unseen setting. Together, these findings provide evidence that VR motion is a viable and sensitive signal of a user's internal state and that it contains much richer information than previously assumed. Our results further demonstrated that pretrained motion foundation models generalize more effectively to previously unseen users than classical machine learning and temporal models, highlighting the value of large-scale motion pretraining for XR inference tasks.

\subsection{Implications for Theory}

% The primary theoretical contribution of this work is demonstrating that the "ceiling" for what motion can reveal is higher than intuitive "folk theories" suggest. While previous research has linked movement to broad emotional traits like depression, we show that motion kinematics also encodes transient cognitive states such as confusion and hesitation, which lack clear, overt physical correlates.

% Furthermore, the proposed VRMotionBERT utilize a VR-native adapter, bridging the gap between sparse VR motion and motion foundation models pretrained on full-body datasets without full-body reconstruction. VRMotionBERT has shown notable performance in data efficiency and cross-generalization. This offers a new methodological pathway for the XR community in motion analysis. By showing that motion foundation models pretrained on large-scale full-body datasets can be adapted effectively to sparse VR signals via a lightweight VR motion adapter, we encourage future research in large-scale motion pretraining for XR inference tasks.

% ----------

The primary theoretical contribution of this work is probing the limits of what can be inferred from motion. Rather than focusing on overt behaviors with clearly related motion patterns, we consider cognitive states that lack clear physical correlates, specifically the distinction between confusion (uncertainty about task interpretation) and hesitation (uncertainty between options). 

Despite the subtlety of this distinction, our results show that these states can be inferred from sparse VR motion alone, achieving 82\% accuracy with a relatively small dataset. This suggests that motion encodes richer information about internal cognitive processes than commonly assumed, extending beyond intuitive “folk theories” of body language.

More broadly, our findings position VR motion telemetry as a window into moment-to-moment cognition, raising new opportunities for internal state modeling in XR while also reframing motion data as a potentially sensitive behavioral signal.

\subsection{Implications for Practice}

Our findings suggest that VR motion telemetry can support the design of cognitively-aware interactive systems, particularly in tasks involving decision-making processes.

In training and educational settings, systems could detect when a user is uncertain about task instructions (confusion) versus when they are deliberating between alternatives (hesitation), and respond differently in each case. For example, confusion may benefit from additional clarification or simplified instructions, whereas hesitation may benefit from comparison aids or highlighting difference between conceptually similar alternatives.

In collaborative or social VR environments, inferred cognitive states could be used to enrich avatar behavior with subtle cues of uncertainty. Rather than relying solely on explicit communication, systems could surface lightweight indicators (e.g., hesitation) to improve realism and immersive experience.

More broadly, interfaces could adapt to a user’s inferred readiness to act. For instance, systems may delay prompts or avoid interrupting users during periods of hesitation, while accelerating interactions when users are in a ready state. 

Importantly, these applications rely only on standard head and hand tracking, suggesting that such capabilities could be integrated into existing consumer VR systems without additional sensing hardware.

% ------------

% \subsection{Implications for Practice }

% These findings establish motion telemetry as a practical behavioral signal for building "cognitive-aware" adaptive VR systems:

% Adaptive Training and Guidance: Systems could detect when a trainee is in a state of confusion or hesitation and provide real-time scaffolding or alternative explanations to facilitate learning.

% Enriched Social Interaction: In collaborative VR environments, user avatars could be augmented with subtle non-verbal cues (e.g., a "hesitation" indicator) to signal uncertainty and improve communication flow between remote users.

% Intelligent Interfaces: Interfaces could modulate their complexity based on inferred user readiness, either accelerating the workflow for prepared users or simplifying the visual field to reduce cognitive load during moments of high confusion.

\subsection{Privacy and Ethics}

The ability to infer internal cognitive states from standard VR motion telemetry raises important privacy issues. Prior work has shown that head and hand motion can be highly identifying at the individual level. Our findings suggest that such motion data may also encode information about users’ moment-to-moment cognitive processes, such as uncertainty and decision-making dynamics.

This combination introduces a new class of privacy risks: motion data, often treated as low-level behavioral signals, may implicitly reveal sensitive information about how users think, not just how they act. It's worth mentioning that such inferences can be made passively in the background, without requiring explicit input from the user.

These considerations highlight the need for careful design of motion-based inference systems. Future deployments should emphasize transparency, clearly communicating what is being inferred and for what purpose, and should ensure that such inferences are performed only with informed user consent. More broadly, our results motivate further research into privacy-preserving methods for motion data, including mechanisms for limiting, obfuscating, or controlling access to sensitive behavioral signals.

% --------------

% As noted in our analysis, the ability to infer internal cognitive states from ubiquitous motion telemetry raises significant privacy concerns. Since motion data is highly identifying at the individual level, combining it with cognitive-state inference could expose sensitive information about a user's thinking process or mental health condition without their explicit awareness. Any future implementation of this technology must prioritize transparency and incorporate strict data governance, ensuring that such inferences are only performed with clear user consent and for the user's benefit.

\subsection{Limitations and Future Work}

This study is limited by its controlled lab environment,  which may not fully represent the diversity of real-world VR usage. To address this limitation, future work can extend this finding in less controlled, real-world settings, e.g., tracking users' daily motion instead of using carefully designed questions to elicit cognitive states.

Another limitation of this work is that the recruited participants are mostly university students, which may not represent the desired population. Future work could recruit a more diverse group of participants to address this issue.

Additionally, the input signal used in this study is restricted to VR motion only, as decided by our task design. Future work could explore multi-modality input or even utilizing motion-and-language based large models such as MoFM~\cite{jiang2023motiongpt} and MotionGPT~\cite{baharani2025mofm}.

Furthermore, future work can also explore the potential of VR telemetry and try inference on more ambitious tasks, 
e.g., whether latent states such as hunger leave detectable signatures in motion.
% e.g., can we tell if someone is hungry from how they move?

Finally, and most importantly, methods of designing for privacy in VR should continue to be explored. As VR motion data has been shown to be highly identifying and contains meaningful signals of users' cognitive states, we encourage deeper research into protecting motion data privacy and industry guidelines.

\section{Conclusion}
\label{sec:conclusion}

We presented a novel dataset and modeling framework for inferring nuanced cognitive states, specifically confusion, hesitation, and readiness, from standard VR telemetry. Our experiments demonstrate that these subtle decision-related states leave detectable traces in sparse behavioral signals. By introducing a VR-native motion adapter, we provide a reusable pathway for the XR community to leverage large-scale motion foundation models without requiring explicit full-body reconstruction.

In summary, this work confirms that standard VR telemetry contains significantly richer behavioral information than is typically recognized. We also highlight the potential of motion foundation models as an effective solution to cross-user generalization challenges. These findings establish a foundation for responsive, cognitive-aware adaptive VR systems while underscoring the urgent need for robust privacy safeguards to protect users' sensitive behavioral data.

\acknowledgments{The authors acknowledge the use of ChatGPT 5.4, Gemini 3, and Claude 4.6 Sonnet for refining the manuscript's prose based on a human-written draft.
Additionally, the Nano Banana 2 model was utilized to generate the left and right panels of \cref{fig:teaser}, while the middle panel was manually drawn.
% and the use of Nano Banana 2 for generating the left and right part of the \cref{fig:teaser}, excluding the middle.
All other figures and tables were created by the authors without the use of generative AI. The authors have reviewed and edited all generated content and assume full responsibility for its accuracy and integrity.}

% %% \section{Introduction} %for journal use above \firstsection{..} instead
% This template is for papers of VGTC-sponsored conferences which are \emph{\textbf{not}} published in a special issue of TVCG.

% \section{Conclusion}

% \lipsum[1]%

% \section*{Supplemental Materials}
% \label{sec:supplemental_materials}

% Refer to the instructions for this section (\cref{sec:supplement_inst}).
% Below is an example you can follow that includes the actual supplemental material for this template:

% All supplemental materials are available on OSF at \url{https://doi.org/10.17605/OSF.IO/2NBSG}, released under a CC BY 4.0 license.
% In particular, they include (1) Excel files containing the data for and analyses for creating \cref{tab:vis_papers} and \cref{fig:vis_papers}, (2) figure images in multiple formats, and (3) a full version of this paper with all appendices.
% Our other code is intellectual property of a corporation---Starbucks Research---and there is no feasible way to share it publicly.

% \section*{Figure Credits}
% \label{sec:figure_credits}

% Refer to the instructions for this section (\cref{sec:figure_credits_inst}).
% Here are the actual figure credits for this template:

% \Cref{fig:teaser} image credit: Scott Miller / Special to the Vancouver Sun, January 22, 2009, page A6.

% \Cref{fig:vis_papers} is a partial recreation of Fig.\ 1 from \cite{Isenberg:2017:VMC}, which is in the public domain.

% %% if specified like this the section will be committed in review mode
% The authors wish to thank A, B, and C. This work was supported in part by
% a grant from XYZ.}

%\bibliographystyle{abbrv}
\bibliographystyle{abbrv-doi}

\bibliography{template}
\end{document}